\documentclass[aps,pre,11pt,amsmath,amssymb,twocolumn,reprint]{revtex4-1}

\usepackage{amsmath}
\usepackage{graphicx}
\usepackage{verbatim}
\usepackage{color}
\usepackage{hyperref}
\usepackage{subcaption}
\usepackage{nicefrac}
\usepackage{lipsum}
\usepackage{upgreek}

\usepackage[normalem]{ulem}

\newcommand{\q}{{\boldsymbol q}}
\newcommand{\x}{{\boldsymbol x}}
\renewcommand{\r}{\hat{{\boldsymbol r}}}
\newcommand{\p}{{\boldsymbol p}}

\begin{document}

\title{Self-assembly, Buckling \& Density-Invariant Growth of\\Three-dimensional Vascular Networks} 
\author{Julius B. Kirkegaard}
\author{Bjarke F. Nielsen}
\author{Ala Trusina}
\author{Kim Sneppen}

\affiliation{Niels Bohr Institute, University of Copenhagen, 2100 Copenhagen, Denmark}

\date{\today}

\begin{abstract}
The experimental actualisation of organoids modelling organs from brains to pancreases
has revealed that much of the diverse morphologies of organs are emergent properties
of simple intercellular ``\textit{rules}" and not the result of top-down orchestration.
In contrast to other organs, the initial plexus of the vascular system is formed
by aggregation of cells in the process known as vasculogenesis.
Here we study this self-assembling process of blood vessels in three dimensions
through a set of simple rules that align intercellular apical-basal and planar cell polarity.
We demonstrate that a fully connected network of tubes emerges
above a critical initial density of cells.
Through planar cell polarity our model demonstrates convergent extension,
and this polarity furthermore allows for both morphology-maintaining
growth and growth-induced buckling.
We compare this buckling to the special vasculature
of Islets of Langerhans in the pancreas
and suggest that the mechanism behind the vascular density-maintaining
growth of these islets could be the result of growth-induced buckling.
\end{abstract}

\maketitle

\section{Introduction}
Tubes are ubiquitous features of numerous biological systems.
In humans, they form the gastrointestinal tract,
the ductal network of the pancreas,
the fallopian tube, the urinary tract, and so on;
with the most obvious example being the
entire vascular network of blood vessels.
On the relevant time scales of multicellular energy consumption,
diffusion is limited 
to deliver metabolites
over length scales smaller than $\sim \! 100 \, \upmu$m.
Instead, on larger length scales tissue need some form of directed transport \cite{Goldstein2015}.
In vertebrates, this active transport is provided by the beating heart through the vascular network,
which in turn has to branch into every part of the organism to nourish tissue and remove waste.

The development of the vascular network involves mainly two processes:
vasculogenesis and angiogenesis \cite{Udan2013, Cleaver2010}.
During vasculogenesis individual endothelial cells coalesce
and \textit{de-novo} form functional vessels \cite{Davis2002, Hogan2002, Etcoff2002}.
Studies of vasculogenesis \textit{in-vitro} have mainly been restricted to two dimensions \cite{Gamba2003},
but recently three-dimensional vascular organoids have been produced \cite{Chen2017}.
Vasculogenesis results in a randomly connected vascular plexus, 
which is subsequently remodelled by pruning or branching
\cite{Kurz2001, Ochoa-espinosa2015, Dodds2010, Restrepo2006}
to a mature vascular network \textit{e.g.} with a hierarchical tree-like structure.
In angiogenesis the tree-like structure is formed by
branching processes involving either splitting (intussusception)
or sprouting dynamics from already formed blood vessels \cite{Udan2013, Preziosi2006, Kurz2015}.
This remodelling can be guided by blood flow, pressure and vessel wall stresses \cite{Pries2005}.
We shall be interested in modelling blood vessels organoids,
and will thus not consider this latter reorganisation,
which becomes relevant only in connection with 
certain organs (such as a pumping heart).

From a theoretical and computational viewpoint,
the most intriguing feature of vasculogenesis is its
three-dimensional self-assembly of tubular networks.
Of equal importance is whether these self-assembled networks
\textit{percolate} across the tissue,
\textit{i.e.} whether a fully connected network of tubes is formed.
What densities of endothelial cells are needed in three dimensions
to ensure this criteria?
We will additionally be interested in questions of growth.
Once a network is formed, can this network undergo stable growth?
And what are the possible mechanisms for such networks to grow
while maintaining a constant space-to-vessel density?

Understanding blood vessel formation computationally has received much attention \cite{Ambrosi2005}.
Continuum models enable descriptions of density fields of chemotaxing endothelial cells during vasculogenesis \cite{Manoussaki1996, Murray2003, Namy2004, Gamba2003}. Likewise, cellular Potts models \cite{Merks2006} and
models of individual cells \cite{Merks2008} have been employed.
These studies of vasculogenesis have focused mainly on two-dimensional systems.
In this paper we introduce a coarse-grained description of tubes in three dimensions
using a formulation that resolves both features of single cells and full organs (vessel network). In particular, we are able to simulate vascular networks
comprised of up to hundreds of thousands of particles.

\begin{figure}[tb]
\centering
\includegraphics[width=0.45\textwidth]{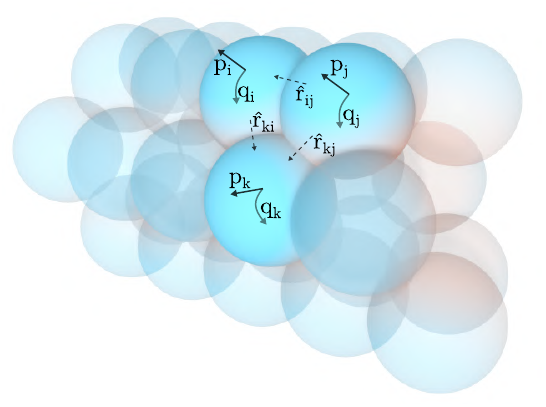}
\caption{The polarity of each particle, which stems from a distribution of proteins
is modelled as vectors. AB-polarity is indicated by $\p$ and PCP by $\q$.
For a tube, as illustrated, $\p$ points away from the tube (distinguishing outside from inside), and $\q$ curls around
the tube (distinguishing along vs. around the tube).
Interactions between cells depend on their orientation between each other $\hat{\r}_{ij}$.
In equations (\ref{eq:s1}-\ref{eq:s3}),
$S_1$ favours $\p_i$ and $\p_j$ parallel and both orthogonal to $\hat{\r}_{ij}$,
$S_2$ favours $\p$'s and $\q$'s orthogonal,
and $S_3$ favours $\q_i$ and $\q_j$ parallel and both orthogonal to $\hat{\r}_{ij}$.
Note that in reality cells can deform based on the polarities, but we 
model them as point particles.
Shape-deformation is captured by collections of multiple particles.
}
\label{fig:illustration}
\end{figure}

\begin{figure}[b]
\centering
\includegraphics{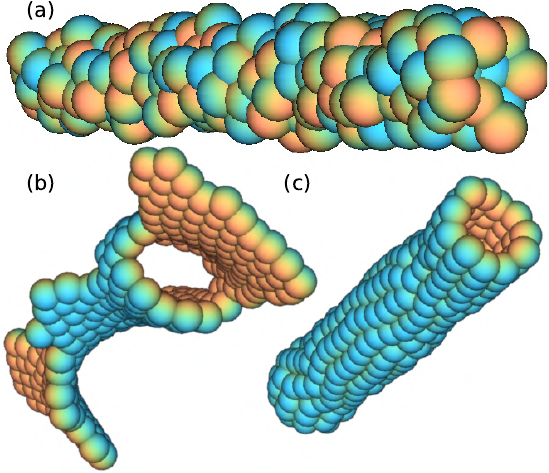}
\caption{Hollowing of a solid tube with self-organising AB-polarity. (a) Initial solid-tube, particles having only spherical interactions ($\lambda_0 = 1$) and random AB-polarity. (b) Self-organising of AB-polarity with $\gamma = 0$. Similar behaviour is observed with $\lambda_1 = 1.0$ and with $\lambda_1=0.5$, $\lambda_2 = 0.42$, $\lambda_3=0.08$ with PCP already organised. (c) Self-organising of AB-polarity with $\lambda_1 = 1.0$ and $\gamma = 5.0$. Tubes are enclosed in simulations,
but cut open for illustration. Colours indicate AB-polarity.}
\label{fig:hollowing}
\end{figure}

\begin{figure*}
\centering
\includegraphics{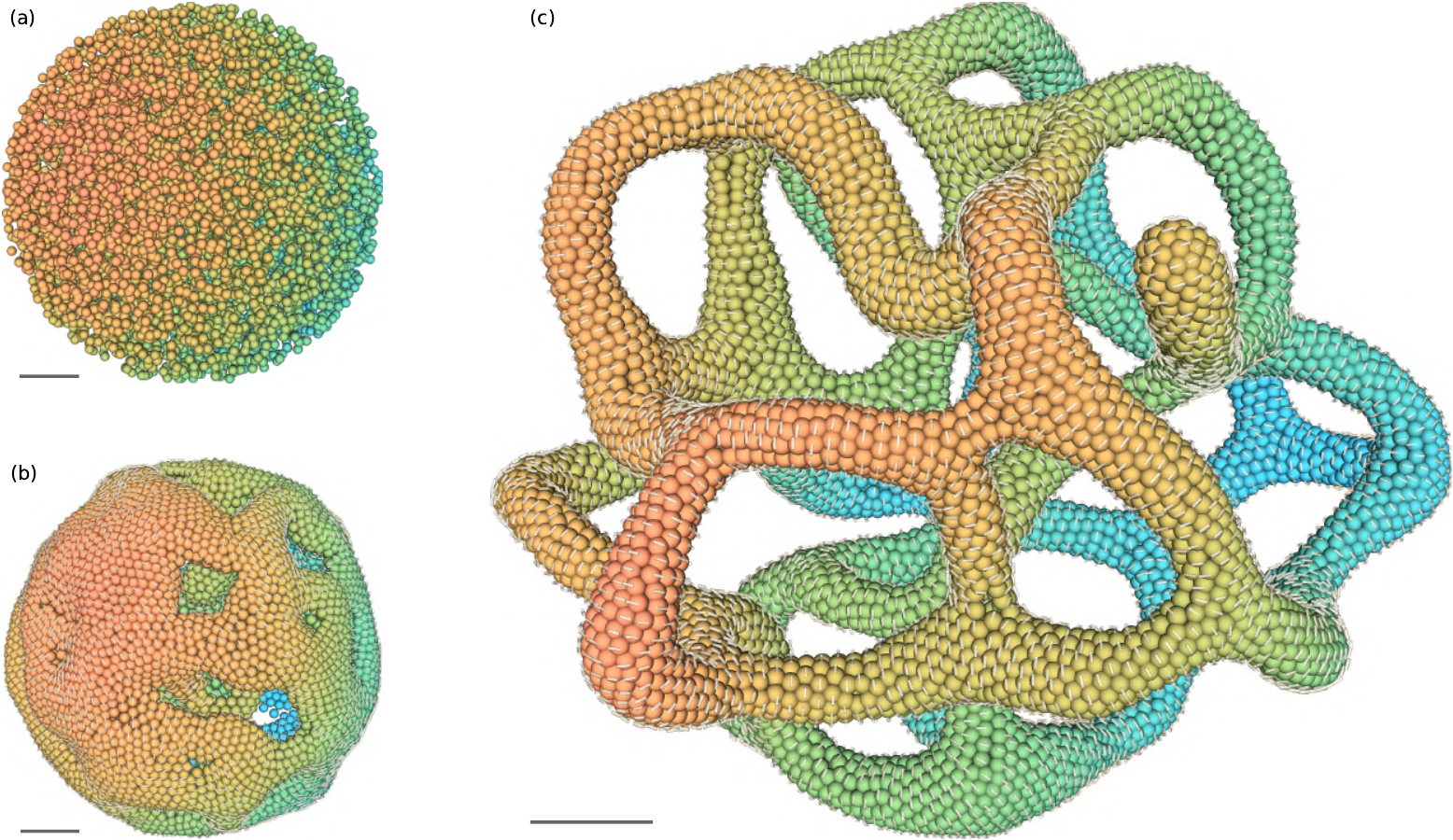}
\caption{Self-assembly of vascular network from $\sim 10^4$ particles initialised uniformly randomly within a sphere.  ($\q$). (a) Initial conditions. (b) Early equilibration of polarisation at $t = 2 \cdot 10^2$, in simulation units. (c) Vascular network formed at $t = 10^4$. Parameters: $\lambda_1 = 0.5$, $\lambda_2 = 0.42$, $\lambda_3 = 0.08$, $\gamma = 5.0$. Scale bar: 5 particle diameters. Lines indicate direction of PCP. Colours for visualisation purposes only.}
\label{fig:selfassembly}
\end{figure*}

Our focus will be on emergent features of the model such as vasculogenesis and
buckling during growth, but we note that our model also has the ability to describe
angiogenetic sprouting \cite{Bjarke} (\textit{budding}) and intussusception,
which on the cell level resembles gastrulation \cite{Nissen2018}.
While lumen formation in blood vessels is its own research field
\cite{Lubarsky2003, Xu2011, Hogan2002, Iruela-Arispe2013},
we introduce a simple mechanism for lumen formation by describing the evolution
of the apical-basal polarity of cells
thus yielding a fully emergent approach to tubulogenesis.

The paper is organised as follows:
In Section II we introduce the methodology and mathematics of the model and demonstrate lumen formation.
Section III is devoted to vasculogenesis and the percolation of the vascular network.
In Section IV we study the growth of vascular network for various parameters of the model and show that both morphology-maintaining and buckling growth patterns can arise.
Lastly, in Section V we describe the vasculature of the Islets of Langerhans in the pancreas,
and describe how their tortuous features could be the result of buckling during growth.
In particular we show that the vascular density during the growth of these
islets could be maintained simply by a buckling mechanism without the need for
angiogenetic processes.

\section{Model, Polarities \\ \& Lumen Formation}
Contrary to 2D models, in three dimensions cell polarity is crucial to model cell sheets.
Our coarse-grained model describes a collection of particles/cells each defined by
their position $\x$, their apical-basal polarity (AB) $\p$, and their 
planar cell polarity (PCP) $\q$,
illustrated in Fig. \ref{fig:illustration}.
Our model is coarse-grained in the sense that a collection of particles
model a cell,
and as such even though each particle is a sphere,
shape deformations are possible in a collection of particles.
In a tube, such as a blood vessel, the AB polarity of cells will define the inside vs.
the outside of the vessel,
while PCP defines the direction around the tube
vs. the direction along the tube.

To model cell behaviour we use a slightly modified version of the model of Ref. \citep{Nissen2018}.
In this model, particles interact pairwise
only if they are line-of-sight Voronoi neighbours
and their mutual potential energy is
\begin{equation}
V_{ij} = \exp(-r_{ij}) - S_{ij} \exp (-r_{ij}/\beta ),
\label{eq:vij}
\end{equation}
for which
\begin{equation}
S_{ij} = \lambda_0 + \lambda_1 S_1^{ij} + \lambda_2 |S_2^{ij}| + \lambda_3 |S_3^{ij}|,
\label{eq:sij}
\end{equation}
where
\begin{align} \label{eq:s1}
&S_1^{ij} = (\p_i \times \r_{ij}) \cdot (\p_j \times \r_{ij}), \\ 
\label{eq:s2} &S_2^{ij} = (\p_i \times \q_i) \cdot (\p_j \times \q_j), \\
&S_3^{ij} = (\q_i \times \r_{ij}) \cdot (\q_j \times \r_{ij}), \label{eq:s3}
\end{align}
and 
\begin{equation}
\r_{ij} = \frac{{\boldsymbol r}_{ij}}{r_{ij}} = \frac{\x_i - \x_j}{|\x_i - \x_j|}.
\end{equation}
We keep $\beta = 5$, which sets the inter-particle spacing to $\simeq 2$ units,
and furthermore enforce $\lambda_0 + \lambda_1 + \lambda_2 + \lambda_3 = 1$ with $\lambda_1 \geq \lambda_3$.
This strikingly simple model can describe a plethora of phenomena related
to polarity-driven morphogenesis in organoids \cite{Nissen2018}.
Naturally, real interactions between polarised cells will be much more complex than
the model used here,
indeed these depend on precise distribution of the surface proteins that make up the polarity of the cells.
The interactions used here can be thought of as the first relevant and symmetry-obeying terms
that give rise to polarity-aligning cells.
The simplicity of the present model is thus agnostic
towards the underlying microscopic details.
Here we introduce a small extension to this model that permits the \textit{de-novo}
formation of tube-like structures.

The dynamics of the model follows from taking all mobilities to be equal. 
Hence,
\begin{align}
\frac{\partial \x_i}{\partial t} = - \frac{\partial V}{\partial \x_i}, && &\frac{\partial \p_i}{\partial t} = - \frac{\partial V}{\partial \p_i}, && \frac{\partial \q_i}{\partial t} = - \frac{\partial V}{\partial \q_i},
\label{eq:dynamics}
\end{align}
with the norms of $\p$ and $\q$ kept at unity.
For this study,
the model was implemented using \textsc{PyTorch} and
ran with \textsc{cuda}-acceleration.

With only spherical interactions, 
\textit{i.e.} $\lambda_0 = 1.0$ a solid tube is
a meta-stable structure of this model as shown in Fig. \ref{fig:hollowing}a.
Lumen-formation corresponds to the formation of AB-polarity, \textit{i.e.} the discrimination of the inside and outside of the tube. 
Various methods for lumen-formation exists, \textit{e.g.} extracellular cord hollowing and lumen ensheathment, or intracellular vacuole fusion \cite{Schuermann2014}.
While the specifics of these mechanism vary, they all establish the AB-polarity 
of the tubes.
If we turn on AB-polarity in the present model, that is we let $\lambda_0 \rightarrow 0.0$
and $\lambda_1 \rightarrow 1.0$, the solid structure tube of Fig. \ref{fig:hollowing}a
opens up into a sheet-like structure as shown in Fig. \ref{fig:hollowing}b.
This behaviour occurs because of the random initialisation of the AB-polarity
(as illustrated in Fig. \ref{fig:hollowing}a).

\begin{figure}[tb]
\centering
\includegraphics{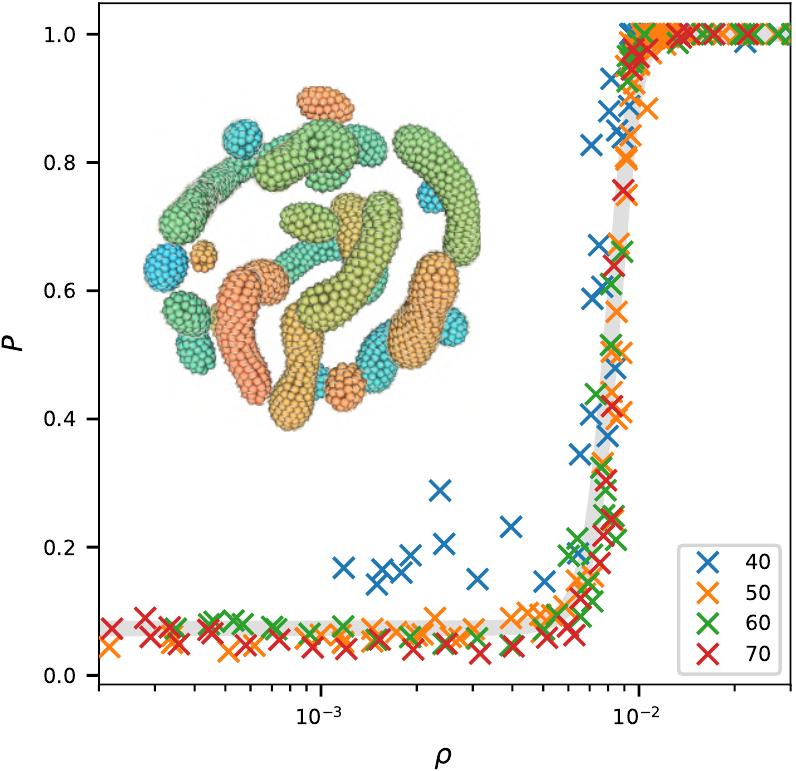}
\caption{Percolation of vascular networks. Graphs show the probability $P$ for particles to belong to the largest cluster as a function of the initial density $\rho = n / (\nicefrac{4}{3} \, \pi \, r^3)$, where $n$ is the number of particles and $r$ is the
radius of the initialisation sphere, its value indicated by the legend.
The critical density is found to be $\rho_c \sim 8.2 \cdot 10^{-3}$.
Inset shows the small clusters for $n = 3,\!300$ and $r = 50$. Parameters: $\lambda_0 = 0.0$, $\lambda_1 = 0.5$, $\lambda_2 = 0.45$, $\lambda_3 = 0.05$, $\gamma = 5.0$.
The critical percolation density $\rho_c$ depends on $\lambda_3$,
since thinner structures percolate more easily.}
\label{fig:percolation}
\end{figure}

To allow AB-polarity to form properly we 
introduce the potential
\begin{equation}
V_i = \gamma \sum_j f(r_{ij}) \, \p_i \cdot \r_{ij}
\label{eq:greens}
\end{equation}
where $f(r) \simeq e^{-r^2/2 \ell^2}$,
such that the total potential is
\begin{equation}
V = \sum_{ij} V_{ij} + \sum_i V_i.
\end{equation}
This potential aligns AB-polarity against 
local areas of high density,
in correspondence to experiments 
suggesting cell-cell contact
directs AB-polarity \cite{Strilic2009}.
It can also be thought of as alignment along a gradient field $c$,
\begin{equation}
V_i = \gamma \, \p_i \cdot \nabla c,
\end{equation}
where $c$ is a molecular, diffusing field of particles
nucleated at cell locations,
\begin{equation}
D \nabla^2 c = \kappa \, c - \sum_i \delta(x_i).
\label{eq:pde}
\end{equation}
This formulation assumes the existence of such a molecular field.
Although many molecular gradients are set up during blood vessel growth,
such as VEGF which elongates and reorganises cells \cite{Etcoff2002},
it is unclear if these interact with 
and orient the polarity of cells.
It is thus easier to think of the interaction as a direct cell-cell interaction
as described by Eq. \eqref{eq:greens}.

\begin{figure}[tb]
\centering
\includegraphics{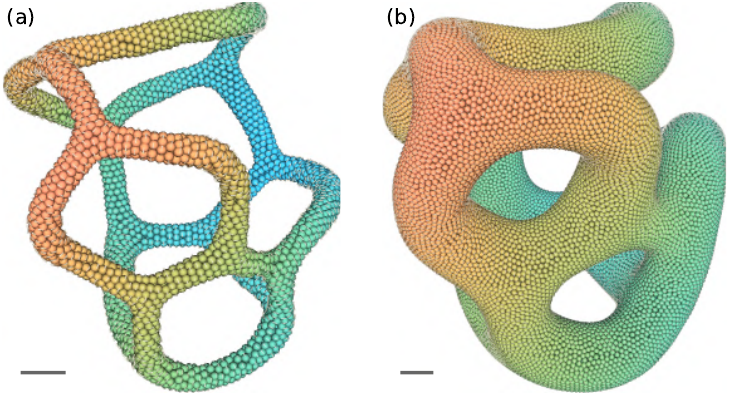}
\caption{Growth of vascular network with $\lambda_3 = 0$. (a) Steady structure formed of $n=3,\!500$ particles with $\lambda_3 = 0.08$. (b) After cell-division to $n=25,\!000$ particles with $\lambda_3 = 0$. Remaining parameters: $\lambda_0 = 0.0$, $\lambda_1 = 0.5$, $ \lambda_2 = 0.42$, $\gamma = 5.0$. Scale bar: 5 particle diameters.}
\label{fig:growthnolam3}
\end{figure}

With $\gamma > 0$ lumen formations occurs and the solid tube becomes hollow and fully enclosed, as shown in Fig. \ref{fig:hollowing}c.
While network formation and lumen formation in reality are separate processes in this
study we will consider the simplified system of them occurring simultaneously.
While $\gamma > 0 $ ensures enclosed structures,
PCP with $\lambda_3 > 0$ is needed to
control the tube thickness.
That is, $\lambda_3>0$ creates a preference for length-wise alignment of particles,
and thus establishes \textit{convergent extension},
which in turn happens 
through cell intercalation events.
Mathemtically, the only difference between AB ($\p$) and PCP ($\q$) is from
the fact that $\lambda_1 > \lambda_3$.
The $\lambda_2$ term keeps $\p$ and $\q$ approximately orthogonal,
and the magnitude of $\lambda_3$ thus determines how much AB alignment
is favoured over PCP alignment,
which in turn controls the thickness of the tubes since thicker
tubes will have better aligned PCP.

In Eq. \eqref{eq:sij} we include vectorial interactions of AB-polarity ($S_1$),
but only nematic interactions of PCP ($S_2$, $S_3$),
since we do not want to impose a 
handedness to the vascular tubes.
At branch points of the vascular network, there must be defects in PCP alignments,
since, in a similar fashion to how you cannot perfectly comb the hair on a sphere,
you cannot have a smooth surface vector field at a tube branch point.
Taking the absolute value in Eq. \eqref{eq:sij} turns vectorial $-1$ charge defects
into two \nicefrac{--1}{2} defects,
which establishes symmetric branch points (see SI).

Finally, we note that we only model the endothelial cells themselves;
in the jargon of active matter research our model is ``dry''.
In organoid experiments there will naturally also be culture media, extracellular matrix,
pericytes, etc., present,
and the system perhaps embedded in \textit{e.g.} matrigel and collagen \cite{Wimmer2019}.
These components mitigate their own interactions between one another and with the cells,
and could be explicitly modelled in a similar manner as our cell-cell interactions.
Such interactions would complicate our model a lot and make interpretations harder,
but one should keep in mind that the parameters we use effectively include these interactions
and are not due solely to pure cell-cell interactions.
For instance, the effects of viscosity of the culture medium would effectively introduce
mobilities in Eq. \eqref{eq:dynamics}.
Likewise, effects of shaking could be modelled effectively by including external noise
in Eq. \eqref{eq:dynamics}.
We have tested such effects and our results remain qualitatively unchanged.

\begin{figure*}[t]
\centering
\includegraphics{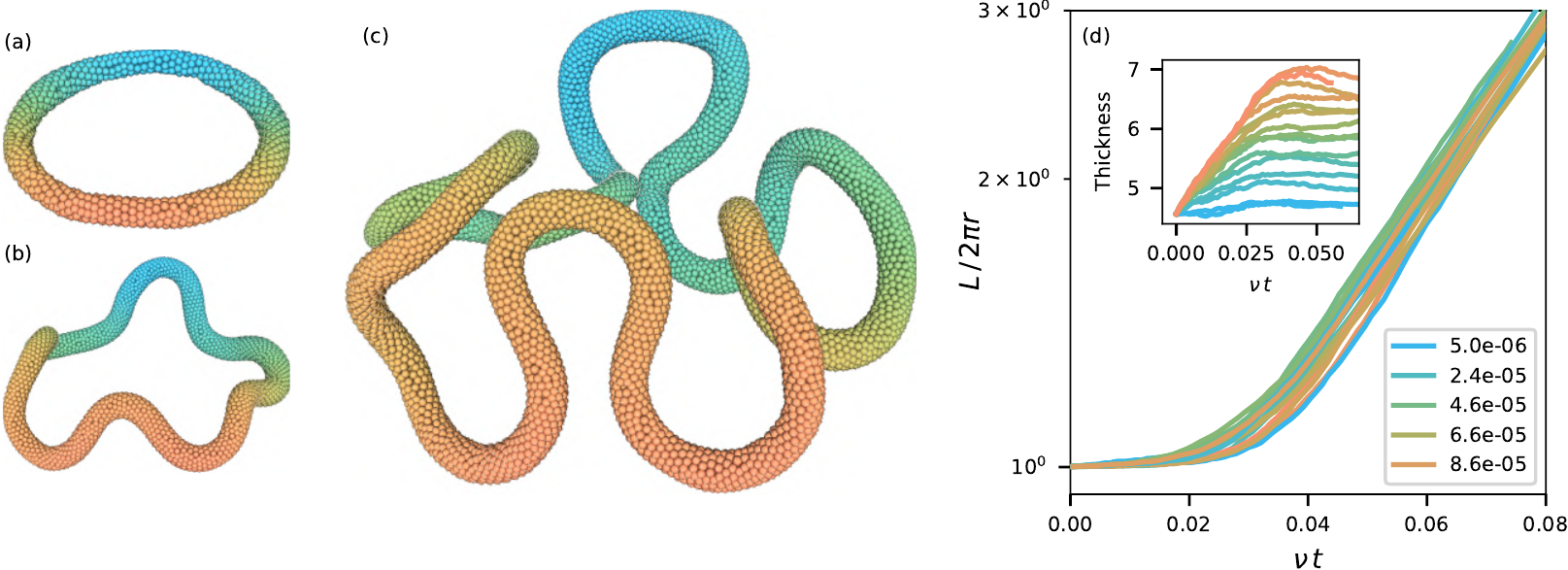}
\caption{Buckling of vessels during cells growth with $\lambda_3>0$. (a) Initial condition of a torus of $1,\!000$ cells. (b-c) Buckling during growth. (d) Buckling measured as vessel contour length $L$ over effective radius $r$ as a function of time re-scaled by division rate $\nu$, which is shown by colour and its value given in the legend. $\nu t \sim 0.03$ corresponding to $\sim 2,\!100$ cells. Inset shows tube thickness ($N/L$) during buckling.}
\label{fig:buckle}
\end{figure*}

\begin{figure}[tb]
\centering
\includegraphics{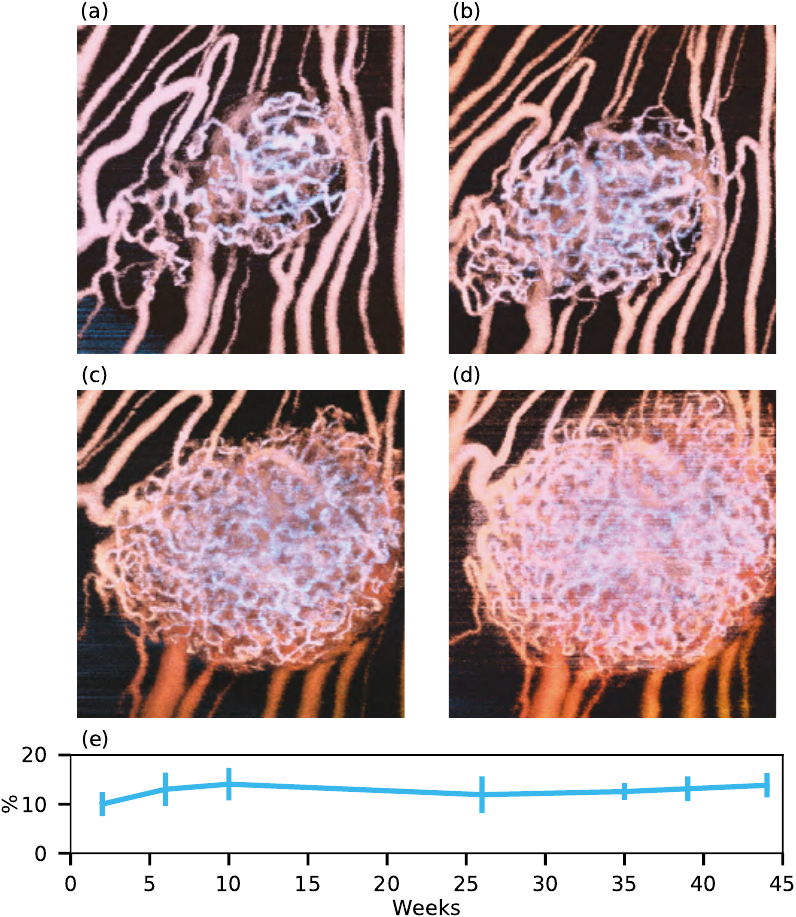}
\caption{Growth of vasculature of islets of Langerhans over 44 weeks.
Images show the vasculature at week 2(a), 6(b), 35(c), and 44(d).
The islet is growing, but its vascular density remains approximately constant,
as shown in (e). Y-axis shows the percentage of vascular volume of the total islet volume.
During these 44 weeks the islet more than triples its volume.
Data and images from Ref. \cite{Berclaz2016}. 
}
\label{fig:islet}
\end{figure}

\section{Vasculogenesis}
During vasculogenesis blood vessels form from aggregating endothelial cells \cite{Udan2013}.
Fig. \ref{fig:selfassembly} shows the self-assembly of three-dimensional vessels in our model.
From an initial random distribution (a) the cells start aggregating (b) and form a tubular network (c).
In Fig. \ref{fig:selfassembly} cells are initially sampled from a uniform distribution within a sphere, but any distribution works.

Naturally, a major concern in vasculogenesis is to form a network of blood vessels
that is fully connected.
It has previously been shown how this \textit{percolation} condition,
\textit{i.e.} whether all particles connect to one another,
depends on the density of endothelial cells \cite{Gamba2003}.
In our model, cells have a preferred distance to one another and 
can attract over long distances.
At first glance therefore it seems that initial density might not be an important quantity.
However, particles only attract if their polarisations match and as soon
as vessel structures have formed, enclosed vessels will not attract one another,
since two vessels nearing each other will have opposing AB polarity on their adjacent surfaces.
Because of this polarisation, the vasculogenesis process in our model is also density dependent.

The density dependent percolation behaviour is visualised in Fig. \ref{fig:percolation},
which shows the probability for a particle to be part of the largest cluster $P$
as a function of the initial density $\rho$.
This is shown for various initial radii, or in other words, for various
number of particles ranging from $\sim 250$ to $\sim 30,\!000$.
As is clear, the vessel network percolates at around $\rho_c \sim 8.2 \cdot 10^{-3}$,
\textit{i.e.} at an initial length scale of $\rho_c^{\nicefrac{-1}{3}} \sim 5$ ---
the same order of magnitude as the inter-particle spacing $= 2$.
Fig. \ref{fig:percolation} also shows some finite-size effects,
since for small number of particles even far below the transition point,
the largest cluster, albeit small, will constitute a significant fraction of the whole system.

\section{Growth \& Buckling}
As organisms grow, so need their network of blood vessels.
The vascular system needs to grow in two distinct ways:
first, blood vessels need to increase their diameter in order to deliver increased amounts of blood.
However, as vessels grow, their surface area to volume fraction decreases
and so their effectiveness.
Thus they also need to grow their network structure to maintain a space-filling network with small diameter vessels, capillaries, at the `leafs' of the network \cite{Gb1997}.
This latter version of growth is called angiogenesis and, as mentioned, is not the focus of our study.
In this section we introduce growth of the blood vessel
and consider the effect of PCP strength $\lambda_3$,
which creates a preference for growth in tube length rather than in tube diameter.
The next section will demonstrate a less considered alternative to space-filling growth 
exploiting the buckling phenomenon demonstrated in this section.

\begin{figure*}[t]
\centering
\includegraphics{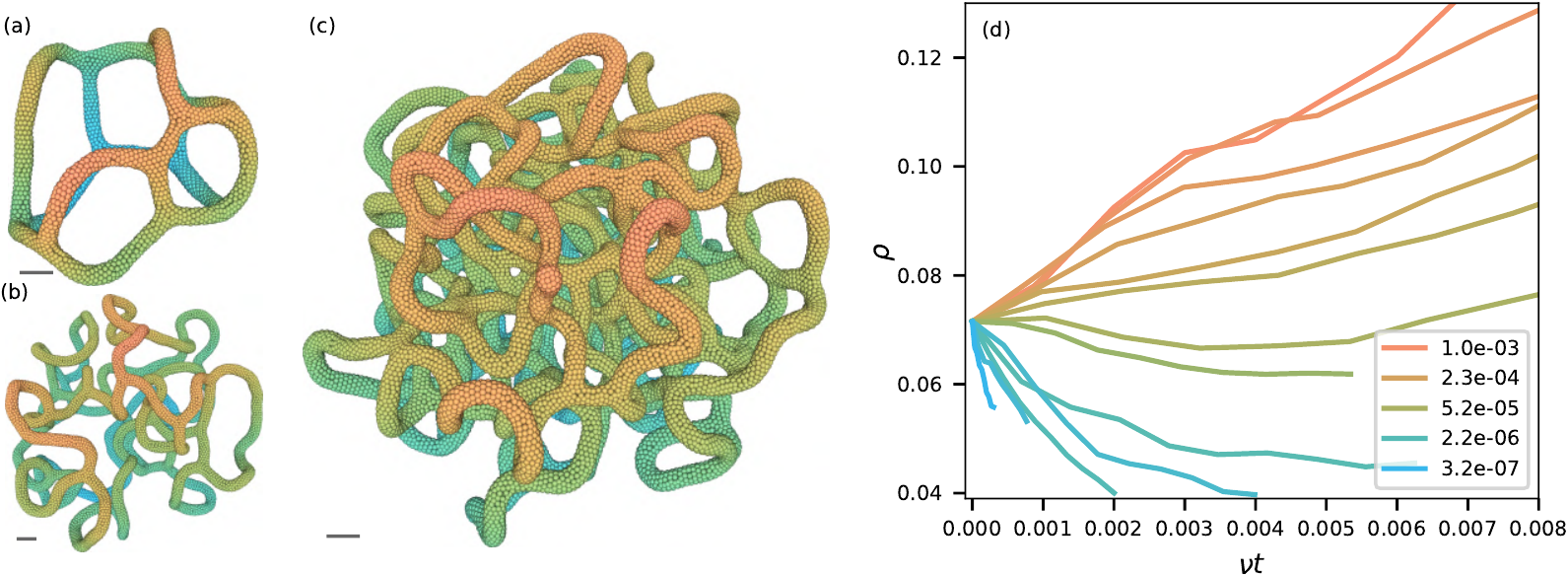}
\caption{Vasculature growth. (a) Initial meta-stable structure of $3,\!500$ particles.
(b) Structure at $\sim 15,\!000$ particles under $\nu \approx 2.5 \cdot 10^{-6}$ growth.
(c) Growth with $\nu \approx 5.0 \cdot 10^{-5}$ having reached the same effective radius as (b),
but at $\sim 45,\!000$ particles and thus at a much higher density.
(d) Vascular density $\rho$ of the structure as a function of time (rescaled by division rate $\nu$, as given in legend). Scale bar: 5 particle diameters.}
\label{fig:density}
\end{figure*}

First we demonstrate that growth of vessel diameters
follows naturally when $\lambda_3 = 0$.
Cell division is implemented as a Poisson process in the sense that
each cell has a constant rate of division $\nu$.
When a cell divides, a new cell is created with the same polarities $\p$ and $\q$,
but placed at a random position next to its mother cell
in the plane orthogonal to $\p$,
meaning that cells divide 
within the cell sheet.

Fig. \ref{fig:growthnolam3} shows the result of growth dynamics with $\lambda_3 = 0$.
Fig. \ref{fig:growthnolam3}a is the steady-state result of
a self-assembly with $\lambda_3 = 0.08$. We then let $\lambda_3 \rightarrow 0$ and $\nu \rightarrow 3 \cdot 10^{-5}$.
Fig. \ref{fig:growthnolam3}b shows the result of the growth
from 3,500 particles to 25,000 particles.
As evident, the vascular network can grow uniformly under this model.
However, the structure as a whole does not grow much,
and in fact the density (vascular volume to free space) grows as well.
This happens because we only model the cell division of the vascular network.
In reality, the tissue between the vessels, which are cells we are not modelling,
will also be dividing and in turn grow the structure as a whole.

If we instead consider the case of $\lambda_3 > 0$ with cell division,
this leads to completely different growth.
Since $\lambda_3$ induces a preferred diameter,
this creates a tendency to grow more in length than thickness.
Fig. \ref{fig:buckle}(a-c) shows how this leads to a growth-induced buckling instability \cite{Drasdo2000},
\textit{i.e.} growth that does not retain the structure's shape.
The buckling occurs because cell division is faster than the time to relax
shape perturbations on long length scales.

A simple measure of buckling of a growing ring is to compare the curve length $L$
with the structure's effective radius $r$.
If no buckling occur $L/r = 2 \pi$.
This definition works well as long as the structure remains a simple curve.
Fig. \ref{fig:buckle}d shows this buckling as a function
of time for various division rates $\nu$.
Clearly, the buckling occurs even for exceedingly small division rates
at approximately the same value of $\nu t$, \textit{i.e.} $\sim $ the same number of cells.
After the transition, the buckling degree grows exponentially in time, or, in other words,
linearly with the number of cells.
In this regime this can be understood as the structure growing mostly in structure length $L$ and not in effective size $r$.
This continues until one side of the structure meets the other.

The inset of Fig. \ref{fig:buckle}d shows the thickness of tubes
calculated as $N/L$, where $N$ is the number of cells in the structure.
As is clear, a high growth rate leads to thicker tubes as the time scale
for growth severely outpaces relaxation.
The thickness of the tubes grows up until buckling starts occurring,
after which the thickness decreases as there suddenly is room to grow in length
instead of thickness and the thickness stabilises.
The peak in thickness occurs at slightly later $\nu t$ for larger $\nu$.

\section{Islets of Langerhans}
In the pancreas, the so-called Islets of Langerhans are responsible for the production of hormones such as insulin.
While these islets only constitute about 1 \% of the pancreatic volume
they contain about 10 \% of the blood vasculature.
This dense vessel network is needed to provide energy to the islets,
but also for intercellular communication \cite{Ballian2007},
and to measure and regulate the production and injection
of hormones into the blood stream.

The dense vasculature of pancreatic islets is shown in Fig. \ref{fig:islet}.
The figure also shows the growth of these islets over 44 weeks.
A crucial observation is that the islets' vascular density,
despite the overall growth of the islets,
remains almost constant \cite{Berclaz2016}.
Furthermore, 
the thickness of the blood vessels also do not change much over this course
(Fig. \ref{fig:islet}; variations in thickness is linked to age \cite{Almaca2014}).
In other words, the growth of the vessel network
is unlike that of Fig. \ref{fig:growthnolam3}.
Angiogenesis is the canonical explanation for the growth of vascular network structure
and this is indeed a factor for pancreatic islets \cite{Ballian2007}.

What is also clear from Fig. \ref{fig:islet} is the tortuous structure of the vasculature.
This is in contrast to normal vasculature which is regular and structured,
as can \textit{e.g.} be seen in the surrounding vessels in Fig. \ref{fig:islet}.
Considering only angiogenesis as a growth mechanism,
the tortuous structure and the space-filling growth
are independent observations.

Both observations can, however, be simultaneously explained by
growth-induced buckling as we demonstrate in Fig. \ref{fig:density}.
Fig. \ref{fig:density}(b-c) shows the different structures that
can be attained in our model  with different growth rates.
As is clear, growing vessel structures with $\lambda_3 > 0$
leads to very tortuous structures.
Furthermore, as shown in \ref{fig:density}d,
the division rate $\nu$ also determines the vascular density.
With a large division rate, the structure does not have time to
relax under the division-induced stress and the vascular density increases.
And conversely, the vascular density is decreased for small division rates.
Hence, a simple feedback mechanism where proliferation rate inversely depends on density,
can keep vascular density constant during growth.
While this effect would definitely be co-occurring with angiogenesis,
it is intriguing that it simultaneously gives an
explanation for the tortuousness of the vascular network.
Although the AB-polarity aligning parameter $\gamma$ is only used for self-assembly,
during this sort growth $\gamma > 0$ can allow for anastomosis,
the fusing separate blood vessels.

\section{Conclusions}
We have demonstrated three-dimensional vasculogenesis
in a simple model that aligns cell polarities through cell-cell interactions.
The initial self-assembly of enclosed structures is ensured
through the alignment of AB-polarity against cell density.
This is the key driver of lumen formation
and enables the \textit{de-novo} formation of tubes.
This interaction also allows for fusing of vessels (anastomosis).
While PCP is not needed for the formation of enclosed structures,
this polarity ensures thin tubular structures and convergent extension.

The self-assembly of the vascular networks results in
fully connected vessels if the particles are initialised above a critical density,
in accordance with previous two-dimensional experiments \cite{Gamba2003}.
Enclosed, non-connected structures appear at lower densities,
and these do not interact, due to their opposing AB polarities.

Introducing cell proliferation in our model leads to distinct behaviour
depending on the strength $\lambda_3$ of PCP,
which controls the preference of tube diameter.
With $\lambda_3 = 0$, we have shown that uniform growth is possible.
With $\lambda_3 > 0$ the tubes buckle under growth.
While blood vessel buckling is typically associated with high blood pressure \cite{Han2013},
this shows that such behaviour can also stem from cell proliferation.

Considering this buckling mode of growth,
we compared it with the vasculature of Islets of Langerhans,
which shows a large degree of tortuousness.
The vessel density of these pancreatic islets furthermore remains
constant during their growth.
We suggest that a simple explanation for this behaviour is growth-induced
buckling in which cell division rate is coupled to vascular density.
This rate, in turn, may be controlled by negative feedback from blood supply to tissue.

Tortuous blood vessels are in general also found in cancerous tumours \cite{Goel2012}.
Cancer growth is often associated with angiogenesis in nearby tissue, a process that we did not explore here,
but could easily be introduced via local cell shape changes \cite{Bjarke}.
The abnormality of tumour vessels is 
thought to be, in part, due
to over-expression of VEGF-A (Vascular Endothelial Growth Factor A) \cite{Nagy2009}.
We have shown how tortuousness in blood vessels can be linked to growth rate.
This could thus also play a major role in the abnormal morphology of tumour vascularisation.

\subsubsection*{Author Contributions}
JBK conceptualised study, designed model, wrote model code, ran experiments and analysis, and drafted the manuscript.
BFN participated in model discussions, was involved in writing model code, and revised the manuscript.
AT provided biological insight and revised the manuscript.
KS conceptualised study, participated in model and analysis discussions, and revised the manuscript.

\subsubsection*{Data Accessibility}
No data was collected.

\subsubsection*{Competing interests}
We declare we have no competing interests

\subsubsection*{Funding}
This project has received funding from the European Research Council (ERC) under the European Union's Horizon 2020 Research and Innovation Programme, Grant Agreement No. 740704
and  the Danish National Research Foundation, Grant No. DNRF116.

\subsubsection*{Acknowledgements}
We thank Anne Grapin-Botton for helpful discussions.

\end{document}